\documentclass[useAMS,usenatbib]{mn2e}
\pdfoutput=1
\usepackage{natbib}
\usepackage{graphicx}
\usepackage{amssymb}
\usepackage{lineno}

\begin{document}
\newcommand{\MSun}{{M_\odot}}
\newcommand{\LSun}{{L_\odot}}
\newcommand{\Rstar}{{R_\star}}
\newcommand{\calE}{{\cal{E}}}
\newcommand{\calM}{{\cal{M}}}
\newcommand{\calV}{{\cal{V}}}
\newcommand{\calO}{{\cal{O}}}
\newcommand{\calH}{{\cal{H}}}
\newcommand{\calD}{{\cal{D}}}
\newcommand{\calB}{{\cal{B}}}
\newcommand{\calK}{{\cal{K}}}
\newcommand{\labeln}[1]{\label{#1}}
\newcommand{\Lsolar}{L$_{\odot}$}
\newcommand{\farcmin}{\hbox{$.\mkern-4mu^\prime$}}
\newcommand{\farcsec}{\hbox{$.\!\!^{\prime\prime}$}}
\newcommand{\kms}{\rm km\,s^{-1}}
\newcommand{\cc}{\rm cm^{-3}}
\newcommand{\Alfven}{$\rm Alfv\acute{e}n$}
\newcommand{\Vap}{V^\mathrm{P}_\mathrm{A}}
\newcommand{\Vat}{V^\mathrm{T}_\mathrm{A}}
\newcommand{\D}{\partial}
\newcommand{\DD}{\frac}
\newcommand{\TAW}{\tiny{\rm TAW}}
\newcommand{\mm }{\mathrm}
\newcommand{\Bp }{B_\mathrm{p}}
\newcommand{\Bpr }{B_\mathrm{r}}
\newcommand{\Bpz }{B_\mathrm{\theta}}
\newcommand{\Bt }{B_\mathrm{T}}
\newcommand{\Vp }{V_\mathrm{p}}
\newcommand{\Vpr }{V_\mathrm{r}}
\newcommand{\Vpz }{V_\mathrm{\theta}}
\newcommand{\Vt }{V_\mathrm{\varphi}}
\newcommand{\Ti }{T_\mathrm{i}}
\newcommand{\Te }{T_\mathrm{e}}
\newcommand{\rtr }{r_\mathrm{tr}}
\newcommand{\rbl }{r_\mathrm{BL}}
\newcommand{\rtrun }{r_\mathrm{trun}}
\newcommand{\thet }{\theta}
\newcommand{\thetd }{\theta_\mathrm{d}}
\newcommand{\thd }{\theta_d}
\newcommand{\thw }{\theta_W}
\newcommand{\beq}{\begin{equation}}
\newcommand{\eeq}{\end{equation}}
\newcommand{\ben}{\begin{enumerate}}
\newcommand{\een}{\end{enumerate}}
\newcommand{\bit}{\begin{itemize}}
\newcommand{\eit}{\end{itemize}}
\newcommand{\barr}{\begin{array}}
\newcommand{\earr}{\end{array}}
\newcommand{\bc}{\begin{center}}
\newcommand{\ec}{\end{center}}
\newcommand{\DroII}{\overline{\overline{\rm D}}}
\newcommand{\DroI}{{\overline{\rm D}}}
\newcommand{\eps}{\epsilon}
\newcommand{\veps}{\varepsilon}
\newcommand{\vepsdi}{{\cal E}^\mathrm{d}_\mathrm{i}}
\newcommand{\vepsde}{{\cal E}^\mathrm{d}_\mathrm{e}}
\newcommand{\lraS}{\longmapsto}
\newcommand{\lra}{\longrightarrow}
\newcommand{\LRA}{\Longrightarrow}
\newcommand{\Equival}{\Longleftrightarrow}
\newcommand{\DRA}{\Downarrow}
\newcommand{\LLRA}{\Longleftrightarrow}
\newcommand{\diver}{\mbox{\,div}}
\newcommand{\grad}{\mbox{\,grad}}
\newcommand{\cd}{\!\cdot\!}
\newcommand{\Msun}{{\,{\cal M}_{\odot}}}
\newcommand{\Mstar}{{\,{\cal M}_{\star}}}
\newcommand{\Mdot}{{\,\dot{\cal M}}}
\newcommand{\ds}{ds}
\newcommand{\dt}{dt}
\newcommand{\dx}{dx}
\newcommand{\dr}{dr}
\newcommand{\dth}{d\theta}
\newcommand{\dphi}{d\phi}

\newcommand{\pt}{\frac{\partial}{\partial t}}
\newcommand{\pk}{\frac{\partial}{\partial x^k}}
\newcommand{\pj}{\frac{\partial}{\partial x^j}}
\newcommand{\pmu}{\frac{\partial}{\partial x^\mu}}
\newcommand{\pr}{\frac{\partial}{\partial r}}
\newcommand{\pth}{\frac{\partial}{\partial \theta}}
\newcommand{\pR}{\frac{\partial}{\partial R}}
\newcommand{\pZ}{\frac{\partial}{\partial Z}}
\newcommand{\pphi}{\frac{\partial}{\partial \phi}}

\newcommand{\vadve}{v^k-\frac{1}{\alpha}\beta^k}
\newcommand{\vadv}{v_{Adv}^k}
\newcommand{\dv}{\sqrt{-g}}
\newcommand{\fdv}{\frac{1}{\dv}}
\newcommand{\dvr}{{\tilde{\rho}}^2\sin\theta}
\newcommand{\dvt}{{\tilde{\rho}}\sin\theta}
\newcommand{\dvrss}{r^2\sin\theta}
\newcommand{\dvtss}{r\sin\theta}
\newcommand{\dd}{\sqrt{\gamma}}
\newcommand{\ddw}{\tilde{\rho}^2\sin\theta}
\newcommand{\mbh}{M_{BH}}
\newcommand{\dualf}{\!\!\!\!\left.\right.^\ast\!\! F}
\newcommand{\cdt}{\frac{1}{\dv}\pt}
\newcommand{\cdr}{\frac{1}{\dv}\pr}
\newcommand{\cdth}{\frac{1}{\dv}\pth}
\newcommand{\cdk}{\frac{1}{\dv}\pk}
\newcommand{\cdj}{\frac{1}{\dv}\pj}
\newcommand{\rad}{\;r\! a\! d\;}
\newcommand{\half}{\frac{1}{2}}

\title[
 Do massive neutron stars end as invisible dark energy objects?]
{Do massive neutron stars end as invisible dark energy objects?  {}}
{}
\author[Hujeirat,  A.A.]
       {Hujeirat, A.A. \thanks{E-mail:AHujeirat@uni-hd.de} \\
\\
IWR, Universit\"at Heidelberg, 69120 Heidelberg, Germany \\
}
\date{Accepted  ...}

\pagerange{\pageref{firstpage}--\pageref{lastpage}} \pubyear{2002}

\maketitle

\label{firstpage}

\begin{abstract}
Astronomical observations reveal a gap in the mass spectrum of  relativistic objects:
neither black holes nor neutron stars having masses in the range of  2 - 5$\,\MSun$  have ever been observed.\\
Based on the solution of the TOV equation modified to include a universal scalar field $\cal{H},$ we argue that
all moderate and massive neutron stars should end invisible dark energy objects (DEOs).

 Triggered by the  $\cal{H}-$baryonic matter interaction, a  phase transition from normal compressible nuclear matter
  into an incompressible quark-superfluid is shown to occur at roughly $3$ times the nuclear density.
  At the transition front, the scalar field is set to inject energy at the maximum possible rate via a non-local  interaction potential
  $V_\phi = a_0 r^2 + b_0.$
  This energy creates a global confining bag, inside which a sea of  freely moving quarks is formed  in line with the asymptotic freedom
 of quantum chromodynamics.
 The transition front, $r_f,$ creeps from inside-to-outside  to reach the surface
 of the object  on the scale of Gyrs or even shorter, depending on its initial compactness.  Having $r_f$  reached $R_\star,$
 then the total injected dark energy via $V_\phi $ turns NSs  into invisible DEOs.\\
 While this may provide an explanation for the absence of stellar BHs with $M_{BH}\leq 5 \MSun$ and NSs with $M_{NS}\geq 2 \MSun$, it also
suggests that DEOs  might have hidden connection  to dark matter and dark energy in cosmology.\\  \\
\end{abstract}


\section{Turbulent superfluidity in Neutron stars}
The interiors of pulsars and NSs  most likely are made of superfluids governed by triangular lattice of quantized vortices as prescribed by the Onsager-Feynman
equation: $\oint \textbf{v}\cdot \textbf{d} l = \DD{2\pi \hbar}{ m}N. $  $\textbf{v},\textbf{d}l,~ \hbar, m$ here denote the velocity field, the vector of line-element, the reduced Planck constant and the mass of the superfluid particle pair, respectively.\\
Accordingly, the core of the Crab pulsar, should have approximately $N_n=  8.6\times10^{17}$
neutron and $N_p \approx 10^{30}$ proton-vortices (Fig. 1).
Let the evolution of the number density of vortex lines, $n_v$, obey the following  advection-diffusion equation:
\beq
            \DD{\D n_v}{\D t} + \nabla\cdot n_v \textbf{u}_f = \nu_t  \triangle n_v,
\eeq
where $t,\, \textbf{u}_f,~\nu_t$ denote the transport velocity at the cylindrical radius $r=r_f$ and dissipative coefficient in the local frame
of reference, respectively. When $\nu_t=0$, then
the radial component of  $\textbf{u}_f $ in cylindrical coordinates reads: $u^{max}_f\approx-(\dot{\Omega}/{\Omega})~r >0.$ In the case of the Crab;
this implies that approximately $10^6$  neutron  vortices must be expulsed/annihilated each second,  and therefore the object should switch off  after $10^{6}$ up to $10^{13}$ yr,
depending on the underlying mechanism of heat transport \cite[see][and the references therein]{Baym1995, Link2012}.
\begin{figure}
\centering {
\includegraphics*[angle=-0, width=8.15cm]{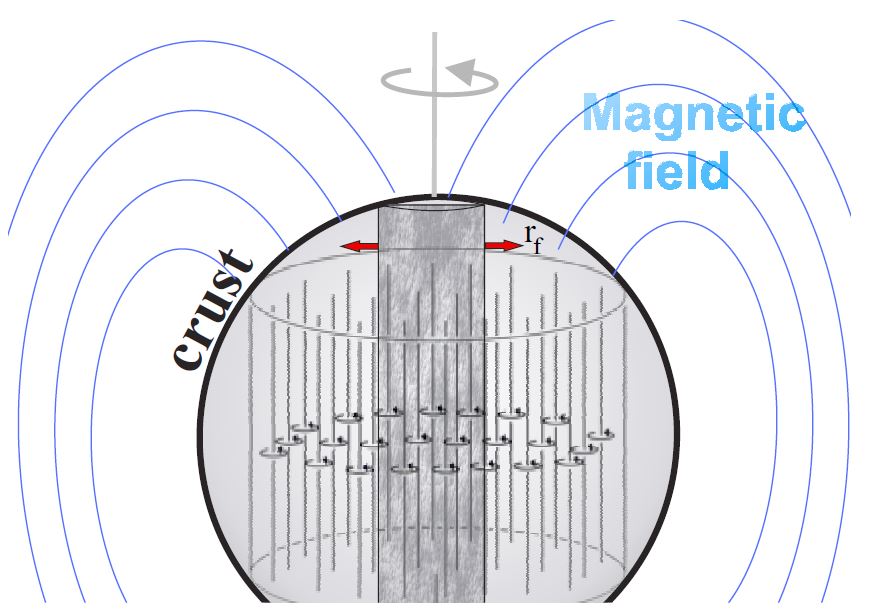}\\
}
\caption{\small  A magnetized neutron star with  a superfluid core threaded  by billions of  vortex lines and magnetic flux tubes. }  \label{NS_Vortices}
\end{figure}
On the other hand, recent numerical calculations of superfluids reveal generation of large amplitude Kelvin waves
that turn superfluids turbulent \cite[see][and the references therein]{Baranghi2008, Baggaley2014, Dix2014}.
It is therefore unlikely that trillions of Kilometer-long neutron and protons-vortices inside pulsars and NSs would behave differently.
 In this case,  $u_f$ should be replaced by  a mean turbulent  velocity $<u_f>^t $ with  $ u^{max}_f$ being an upper
 limit\footnote{The rotational energy associated with the outward-transported vortex lines from the central regions
are turbulently re-distributed in the outer shells and should not necessary suffer a complete annihilation.}.
 As the number of vortex lines decreases with time due to emission of
 magnetic dipole radiation and therefore the separation between them increases non-linearly, it is reasonable to associate a time-dependent turbulent length scale
 $\ell_t(t),$ which covers the two limiting cases: $\ell_t(t=0) = \ell_0 \approx 10^{-3}\,$cm and  $\ell_t(t=\infty)=\ell_{\infty}= R_\star.$
 This yields the geometrical mean   $<\ell_t> = \sqrt{\ell_0 \ell_{\infty}}\approx \calO(10)\,cm.$
  Putting terms together and using $\nu_{tur}= <\ell_t><u_f>^t $  to describe the effective turbulent viscosity, we obtain an upper limit for the global diffusion time scale:
$\tau_{diff} = {R_{NS}^2}/{\nu_{tur}}= \calO(10^9)$ yr.   Similarly,  a comparable time scale for the
Ohmic diffusion in this turbulent medium can be constructed as well.
This is in line with observations, which reveal that most isolated luminous NSs known are younger than $10^9$ yr
  \cite[see][and the references therein]{Espinoza2011}.
Assuming quantized vortices in NSs to obey a triangular lattice distribution, then the very central region would be the first
to be evacuated from vortex lines and all other removable energies that do not contribute significantly
 to the pressure.  Hence the radius of this region, $r_f,$ would creep outwards with an average
velocity:  $\dot{r}_f\sim R_\star/\tau_{diff} \approx10^{-10}$ cm/s. The nuclear matter inside $r_f$  would be
in the lowest possible energy state, which, as  argued here, must be the incompressible quark-superfluid phase.
Having $r_f$ reached $R_\star,$ then the NS turns invisible.\\
  In analogy with normal massive luminous stars, massive and highly compact NSs appear to also switch-off  earlier than their less massive
  counterparts (Hujeirat 2016). For alternative models explaining the above-mentioned mass gap see
  \cite{Belczynski2012}, \cite{Chapline2014}  and the references therein.


\section{The onset of incompressibilitiy}
Modeling the internal structure of cold NSs while constraining their masses and radii  to observations, would
  require their central densities to inevitably be much higher than the nuclear density $-\rho_0$:
a density regime in which all EOSs become rather uncertain and mostly acausal \cite[see][and the references therein]{Hempel2011}.
This however can be  viewed as a consequence of the considerable reduction of the compressibility of the nuclear matter
at $r=0$. To clarify this argument: Assume that the energy density and the pressure at $r=0$ have reached the critical state,
at which the  particle involved communicate with each other at the the maximum
possible speed, e.g. the speed of light. This corresponds to the EOS: $p = \calE.$
In this case, the the chemical potential equation reads:
\[
              \mu =\DD{\D \calE}{\D n} = \DD{P+\calE}{n}= \DD{2\calE}{n}, \mbox{  whose solution is: } \calE = a~ n^2.
\]
Let the fast communicating particles  occupy  the finite central volume  $dV_c = 4\pi \int_0^\epsilon {r^2}dr,$  where $\epsilon$ is an arbitrary small radius.
The particles involved practically form a fluid portion that cannot accept compression anymore, as otherwise  the causality condition would be violated.
The number density here would saturate around $n_{cr}$ and yields a maximum local pressure $P_{cr}=a~ n_{cr}^2.$
With this $n_{cr} $ and $P_{cr}$, the fluid portion inside $dV_c $ is practically  incompressible.
On the other hand,  the energy inside $dV_c$ is uniform and the involved particles share the same energy, i.e.
$\calE = \calE_0 \times n,$ where $\calE_0 = a~n_{cr}.$ But as $\calE/n = const.$ then local pressure $P_L$ must vanish.
\emph{This means that the validity of calculating the pressure from the chemical potential alone $(P = n^2 \DD{\D}{\D n}(\calE/n))$ breaks down.}
 As a consequence, using this formula in this regime would give rise to unrealistically high central energy densities and most likely would violate causality.
Moreover, the regularity condition imposed on the pressure at r=0 enforces the supranuclear dense fluid inside $dV_c$ to also be nearly incompressible.
 To explain this point: since the gradient of the pressure vanishes at $r =0,$ the RHS of the TOV-equation (see Eq. 9) must vanish as well.
This is feasible, if the enclosed mass becomes vanishingly small i.e. $m(r) = 4\pi \int_0^\epsilon  \calE r^2 dr \ll 1$ for $\epsilon \ll 1.$
On such small length scales  and,  in the absence of local or exotic feeding mechanisms,  gravity alone cannot enforce
$ \calE$  to increase faster than  $1/r$ as $r \rightarrow 0.$ Therefore the spatial variation of $ \calE$ inside $dV_c$ remains limited, which means  that $m(r) \approx \calE_0 r^3$  and therefore the formation of the density plateau around $r=0$  becomes inevitable.
 Under these conditions, computing the local pressure $P_L$ from the chemical potential alone would yield an unrealistic $P_L ( \leq 0).$
 The usual adopted strategy to escape this pressure-deficiency is to enforce  an unfounded inward-increase of $\calE$ as $r\rightarrow 0,$ resulting
 therefore in  unreasonably large central densities.\\
\section{The onset of quark-superfluidity}
Another possible solution, which we propose here,  runs as follows:
\begin{itemize}
  \item The nuclear matter at $r=0$ indeed reaches the compressibility limit and can be well-described by  the stiffest EOS $P=\calE.$
  \item  A pure incompressible nuclear matter has a constant chemical potential and therefore the validity of computing the local pressure
                from the chemical potential alone  breaks down.
            In such flows  a  non-local pressure $P_{NL}$ for controlling the dynamics of the nuclear fluid is required.
  \item   The transition from compressible into pure-incompressible fluid phase might be provoked by the onset of a scalar field - matter interaction,
            which become active, once a critical density, $n_{cr},$ is surpassed.  The interaction potential, $V_\phi,$  generates a  non-local negative pressure
            $P_{NL},$ which is capable of supporting the fluid-configuration against its own self-gravity.
  \item   The onset of interaction has a run-away character: $V_\phi$  injects dark energy, which in turn enforces the transition front
              to creep  from inside-to-outside to abruptly terminate  at the surface of the object.
\end{itemize}

 Indeed, beyond $ \rho_0$,  short-range repulsive interactions between particles mediated by the exchange of
 vector mesons most likely will dominate the dynamics of nuclear matter and would enhance the asymptotic
 convergence of the EOSs towards  $P \rightarrow \calE\sim n_b^2$  \cite[see][and the references therein]{Haensel2007,Camenzind2007}.
  The chemical potential here  $\mu (\doteq (\calE +p)/n_b)$ increases linearly with the number density of the baryonic matter $n_b.$
  This regime is classified here as H-State and depicted in red-color  in Fig. (2). \\

Recalling that central densities, $\rho_c,$ in NSs increase with their masses, but upper-bounded by $\rho_c \leq 12,5 \times \rho_0$
 to fit the observed mass function \cite[see][and the references therein]{Lattimer2011}, we conclude that the linear correlation $\mu\sim n_b$
 must terminate at a certain critical density $n_{cr},$ where $\mu$ attains a  global maximum \cite[][]{Baym1976}.

 On the other hand, in an ever expanding universe, the eternal-state of matter should be the one at which the internal energy reaches a
 global minimum in  spacetime (zero-temperature, zero-entropy and where  Gibbs energy per baryon is lowest; henceforth the L-State).
 Taking into account that $\mu(r) e^{\calV(r)} = const.$ inside the object together with the a posteriori results $e^\calV \ll 1$ (see Fig. 5),
  we conclude that  $\mu \sim \calE/n_b = const.$  and therefore the local pressure  $P_L = n_b^2 \DD{\D}{\D n}(\DD{\calE}{n_b}) $
  must vanish as well. In this case  a non-local pressure $P_{NL}$ must be generated in order to oppose
 self-collapse\footnote{An incompressible fluid with $\calE= const.$ has a negative local pressure. Therefore an acausal non-local
   pressure is necessary for stabilizing the configuration.}of NSs into BHs with $M \leq 5\,\MSun.$\\
 If the transition layer between the H and L-states is of finite width in the $n-$space,  then  $d\mu/dn$ here may be positive, negative and/or discontinuous.\\
 However, the case  $d\mu/dn > 0$  should be excluded, as it implies that the eternal state of matter would be more energetic than the H-State, which is
 a contradiction by constrcution.
 Similarly, the case $d\mu/dn < 0$  is forbidden as it would violate energy conservation
  (; $d\mu/dn <0 \Leftrightarrow dP/dn <0 \Leftrightarrow$ adding more particles yields a smaller pressure).
  Moreover, let us  re-write the TOV equation in terms of $\mu$:
  \beq
  \DD{d\mu}{dr} = -\DD{G}{c^2 r^2}(\DD{d\calE}{d\mu})(\DD{d\mu}{dn})(\DD{m+4\pi r^3P}{1- {r_s}/{r}}).
  \eeq
Obviously,   as $\mu >0,$ a negative $d\mu/dn $ would destabilize  the hydrostatic equilibrium, unless external sources are included, e.g. bag energy and/or  external fields.\\
 Therefore, although  a first order phase transition may not be completely excluded, a crossover phase transition
 into an incompressible superfluid phase with $\mu=\mu(n=n_{cr}) = const.$  would be more likely.
 Here, $\mu$ and $P$ on both sides of the transition front  are equal and, with the help of an
   external field, both $(\calE/n)^+ $ and $(\calE/n)^-$ across the front  can be made even continuous (Fig. 2)).\\

 In the present study, the simultaneous occurrence of the onset of $\calH-$baryonic matter interaction with the crossover phase transition
 is necessary in order to generate a non-local pressure with  $\nabla P_{NL}<0$  capable of  opposing compression exerted by the surrounding
 curved spacetime.
  In the regime $\rho > \rho_0,$ such a pressure  may nicely resemble a non-local bag energy of quarks in  the continuum.\\
  In the presence of $\calH$, the chemical potential per particle at $r=0$ would be upper-limited by the energy required for  quark-deconfinement.
   In this case, the corresponding  Gibbs function reads:
 \beq
       f(n) = \DD{\calE_b + \calE_\phi}{n} - 0.939\mbox{ GeV}
 \eeq
 Based on our test calculations,  an interaction potential obeying a power law distribution of the type:
 $V_\phi(r) = a_0 r^2 + b_0$ turns out to be optimal for maximizing the compactness of the compact object, i.e, $ \alpha_s (\doteq r_s/r )\rightarrow 1,$
 where $ r_s$ corresponds to the dynamical Schwarzschild radius (Fig. 5).\\
\begin{figure}
\centering {
\includegraphics*[angle=-0, width=8.15cm]{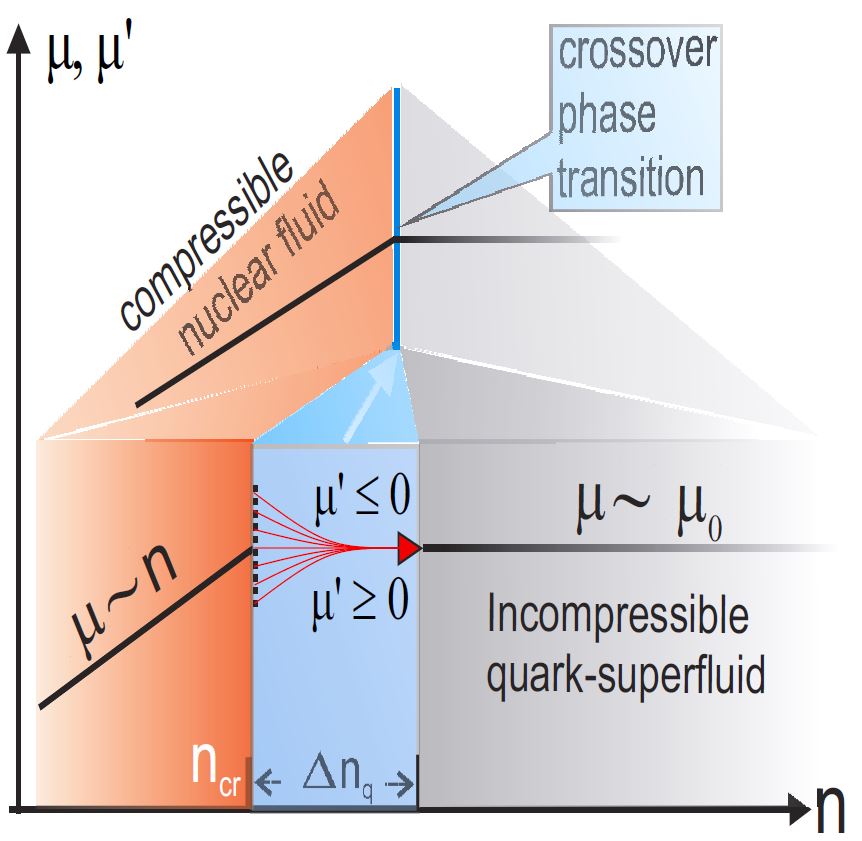}\\
}
\caption{\small  A schematic description of  the chemical potential $\mu,$ $\mu^{'}(\doteq \partial \mu/\partial n)$ versus  the number density $n$ at
the centers of NSs.
 At a critical central density, $n_{cr},$ the universal scalar field $\calH$ is set to provoke a crossover phase transition
 from compressible nuclear fluid into incompressible quark-superfluid. The transition front creeps from inside-to-outside
 to reach the surface of the object on the scale of Gyrs.  }  \label{NS_Vortices}
\end{figure}

Subtituting $\calE_b = a_0 n^2$ and $V_\phi(r)$ in Eq. (3) at $r=0$,  then $f(n)$ reduces to:
 \beq
       f_(n) =  a_0\,n+ \DD{b_0}{n} - 0.9396.
 \eeq
  The Gibbs function here may accept several minima at $n_{min}= (b_0/a_0)^{1/2},$  though $f(n=n_{min})$ doesn't necessary vanish.
  However $f(n)> 0$ and $f(n)<0$ should be excluded, as they are energetically unfavorable for smooth crossover phase transitions to occur. \\
  On the other hand, by varying $a_0\mbox{ and }b_0,$ a set of true minima could be found.
  One way to constrain $b_0$ is to relate it to the canonical energy scale characterizing the effective coupling of quarks, i.e.
  $b_0=0.221$ \cite[see][and the references therein]{Bethke2007}. Indeed,  as shown in Fig (3),   $f(n)$ attains a zero-minimum at $n \approx 3~ n_0$
  for $a_0=1.0.$  \\
The question to be addressed here is whether  the above-mentioned localized analysis would apply  for the whole object as well?\\
 Indeed, the injected dark energy, $E_\phi (= 4\pi \int \calE_\phi r^2 dr)$ via  $V_\phi(r)$ enforces the spacetime embedding the whole object to be
 increasingly curved, thereby maximizing the compression of the fluid in front of $r_f$ up to the critical limit and sets $r_f$ into  an outward motion.
 The enclosed dark energy $E_\phi$ via  $V_\phi(r)$  grows with radius as $r_f^5,$  i.e. faster than the growth of the baryonic mass, thereby enabling the object to reach a  maximum compactness precisely at $r=R_\star.$
 Note that the cases with $E_\phi > r_f^5$ and $E_\phi < r_f^5$ should be excluded. In the former case, the resulting objects must have collapsed
  into BHs with $M< 5 \,\MSun,$ which have not been observed. The latter case is not supported by observation either as the surfaces of  these
  massive NSs  would continue to be dominated by a normal luminous matter.\\
  Behind $r_f,$ a sea of  freely moving quarks is formed, though  globally confined by the strongly curved spacetime surrounding the object,
 which acts as a global confining bag for the quarks.
 Note that, unlike the constant bag energy model of quarks, where the enclosed deconfinement energy scales linearly with the number of
 3-quarks flavors $A$, the injected dark energy in the present model scales as $A^{5/3}.$ This extra-energy may be viewed as a
 mechanism for further enhancing the gluon like-field embedding the  quark-continuum.

 We may examine the conditions of coupling of particles in this pure quark-sea  by setting $\Lambda = b_0$  and taking $N=3$  to be the number of quark flavors in the effective quark-gluon coupling constant:
 \beq
 \alpha_q =  \DD{\pi}{9} \DD{1}{ln(Q^2/\Lambda^2)}.
 \eeq
 Relating $Q$ to Fermi momentum and use $n=n_{cr}=3\, n_0 $ to infer the  Fermi wave number $k_F,$  we obtain $\alpha_q \approx 0.199.$

 However, noting that the sea of quarks is incompressible in which communication between particles is mediated  with the speed of light, we conclude that the value of $\alpha_q$ should attain its true minimum, which is expected to be much smaller than $0.199.$  Nevertheless, the present value of $\alpha_q$ still ensures that quarks are in the safe energy regime, where they move almost freely in line with the asymptotic freedom of quantum chromodynamics -QCD \cite[see][and the references therein]{Bethke2007}. \\

\begin{figure}
\centering {
\includegraphics*[angle=-0, width=8.15cm]{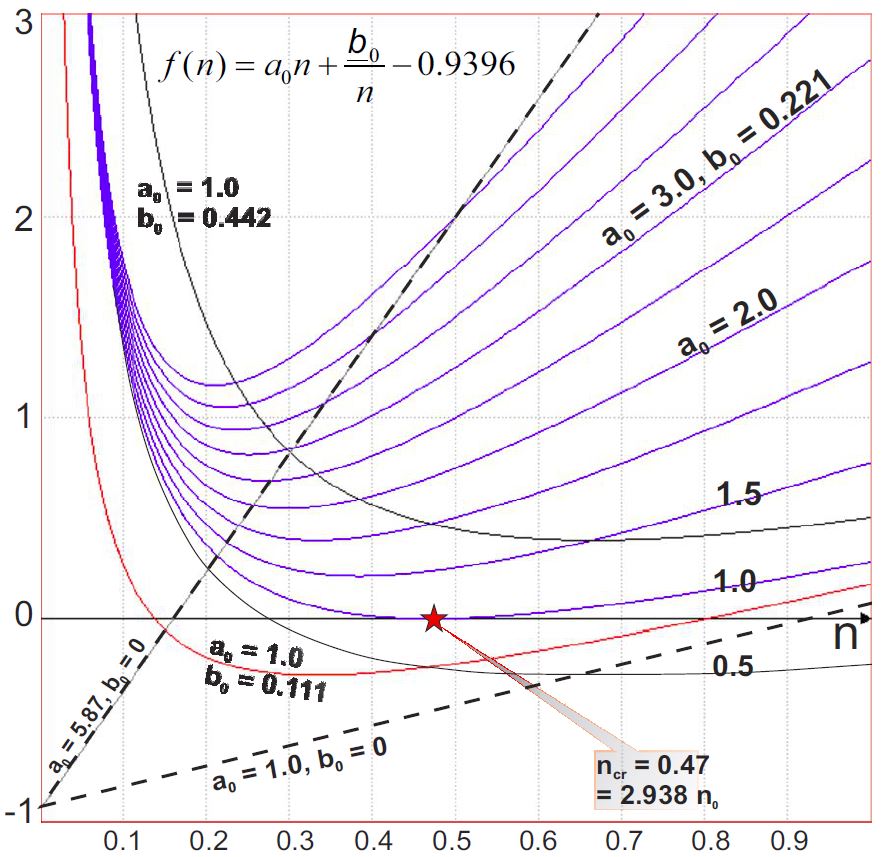}\\
}
\caption{\small  The modified Gibbs function $f(n)$ versus  baryonic number density n (in units of $n_0$) is shown for various values of $a_0$ and $b_0.$
Obviously, $a_0= 1 $ and $b_0=0.221$ appear to be the most appropriate parameters that are compatible  with QCD. The value $b_0=0.221$ corresponds
 to the canonical energy scale characterizing the effective coupling of quarks inside individual hadrons. The Gibbs function $f(n)$ here attains
  a zero-minimum at $n=2.938\, n_0,$  at which  $\calH$ is set to provoke a phase transition into the INQSF state. }  \label{NS_Vortices}
\end{figure}

 \section{Governing equations and solution  method}
 Our investigation here is based on numerical solving the TOV equation modified to include scalar fields $(\cal{H}).$  The modified stress energy tensor  reads:
\beq
 T^{mod}_{\mu\nu} = T^0_{\mu\nu} + T^\phi_{\mu\nu}.
\eeq
The superscripts "0" and "$\phi$" correspond to baryonic and scalar field tensors:
\[
T^0_{\mu\nu} = - P^0 g_{\mu\nu} + (P^0 + \calE^0)U_\mu U_\nu \textrm{ ~~and ~~ }
\]
\beq
T^\phi_{\mu\nu} = (\D_\mu \phi)(\D_\nu \phi) -g_{\mu\nu}[\half (\D_\sigma \phi)(\D^\sigma \phi)-V(\phi)].
\eeq
$U_\mu$ here is the 4-velocity,  the subindices $\mu,~\nu \textrm{ ~  run from 0 to 3 and }  g_{\mu\nu } $\textrm{is a background metric of the form:}
\beq
 g_{\mu\nu} = e^{2\calV}dt^2 - e^{2\lambda}dr^2 - r^2 d\theta^2 - r^2 sin^2\theta d^2\varphi^2,
\eeq
where $\calV,~\lambda$ are functions of the radius.\\
Assuming the configuration to be in hydrostatic equilibrium, then  the GR field equations, $G_{\mu\nu} = - 8\pi G T_{\mu\nu}$  reduce into  the generalized TOV equations:
\beq
\DD{dP}{dr} = - \DD{G}{c^4 r^2} [\calE + P]{[ m(r) + 4\pi r^3 P]}/{(1 - r_s/r)},
\eeq
where ${m(r)} = 4\pi \int \calE r^2 \,dr $  is the total enclosed mass:
$\calE = \calE^0 + \calE^\phi,~ P = P^0 + P^ \phi, $  and where $\calE^\phi = \half \dot{\phi}^2 + V(\phi) + \half (\nabla\phi)^2,$
$ P^\phi = \half \dot{\phi}^2 - V(\phi) - \DD{1}{6} (\nabla\phi)^2.$\\
$V(\phi)$ here denotes the interaction potential of the scalar field with the baryonic matter, i.e., the rate at which dark energy
is injected into the system and $\dot{\phi}$ is the time-derivative of $\phi.$ \\

\begin{figure}
\centering {\hspace*{-0.75cm}
\includegraphics*[angle=-0, width=8.15cm]{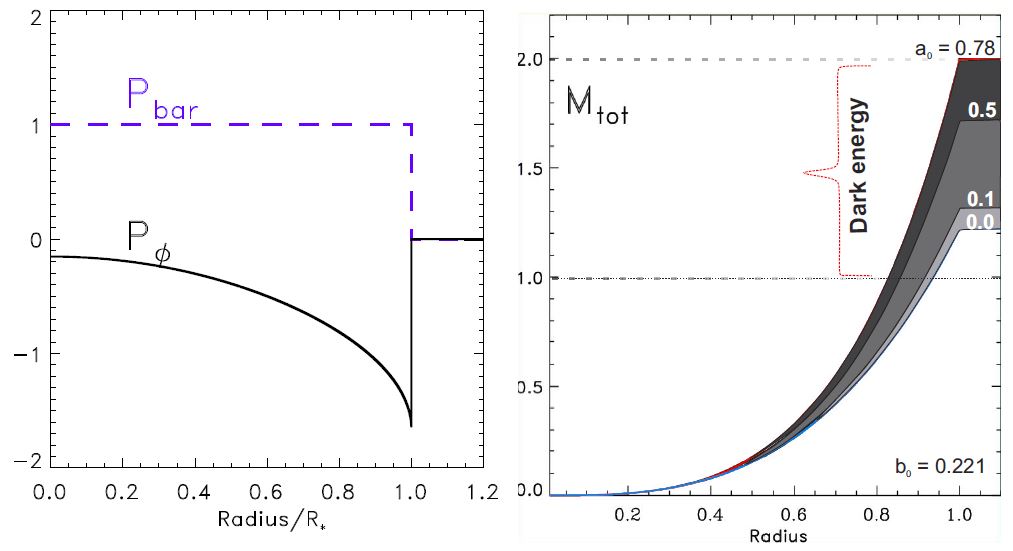}
}
\caption{\small  The radial distributions of the baryonic pressure ($P_{bar}$) and negative pressure  ($P_\phi$)  inside
an incompressible quark-superfluid core (left). The enclosed mass of the baryonic matter and the gradual mass-enhancement
due to dark energy  is shown for different values of  $a_0$ (right).
}  \label{DarEnergy_grr}
\end{figure}
Our reference object is a NS with $1.44~\calM_\odot $  with a radius $R_\star = 2\times R_S,$  where $R_S$ is the Schwarzschild radius.
$\phi$ is assumed to be spatially and temporarily constant, whereas $V_\phi$ is set to obey the power-law distribution:
$V_\phi = a_0\, r^\Gamma +b_0.$  $a_0$ and $b_0$ are constant parameters that are chosen so to fulfill the
a posteriori requirement: $R_\star= R_S + \epsilon,$ for $\epsilon \ll 1.$ In most of the cases considered here,  $b_0$ is set to
be identical to the canonical energy scale at which mometum transfer between quarks saturates, i.e.,  $b_0=0.221$GeV.
The  fluid in the post transition phase is governed by the EOS: $P^0 = \calE^0 = \rho_{cr}c^2= const.$\\
For a given central density, the  solution procedure adopted here is based  on integrating the equations for the pressure, enclosed mass and
pseudo-gravitational potential from inside-to-outside,  using either the first order Euler or fourth order Runge-Kutte integration methods.

\begin{figure}
\centering {\hspace*{-0.75cm}
\includegraphics*[angle=-0, width=8.15cm]{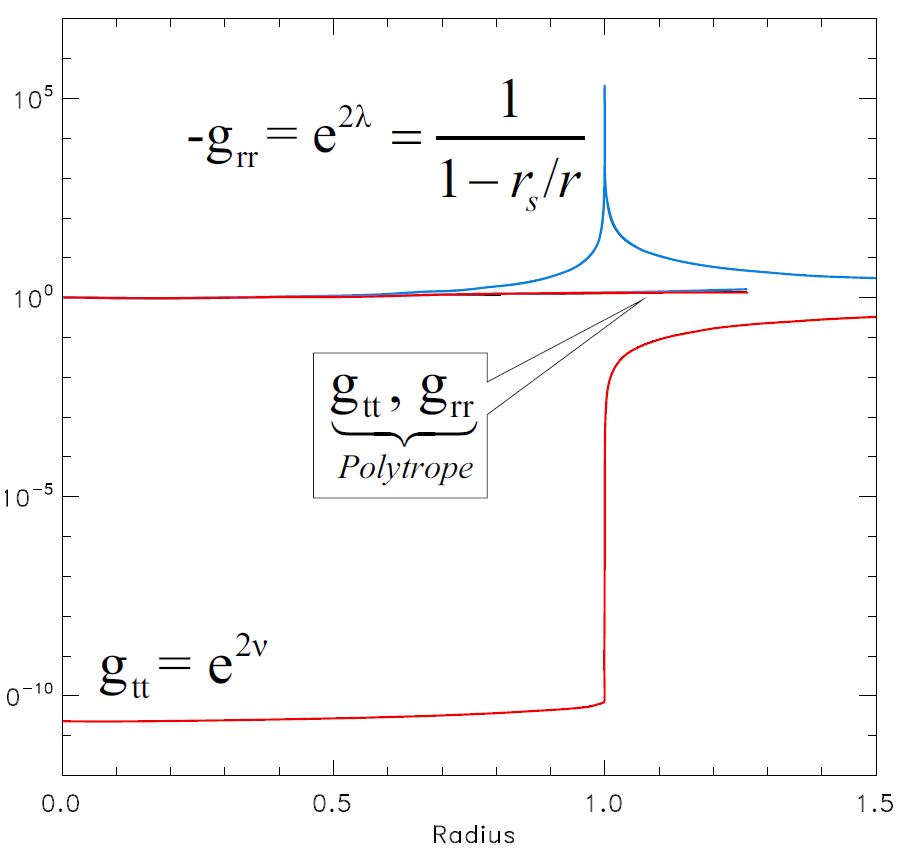}\\
\includegraphics*[angle=-0, width=8.15cm]{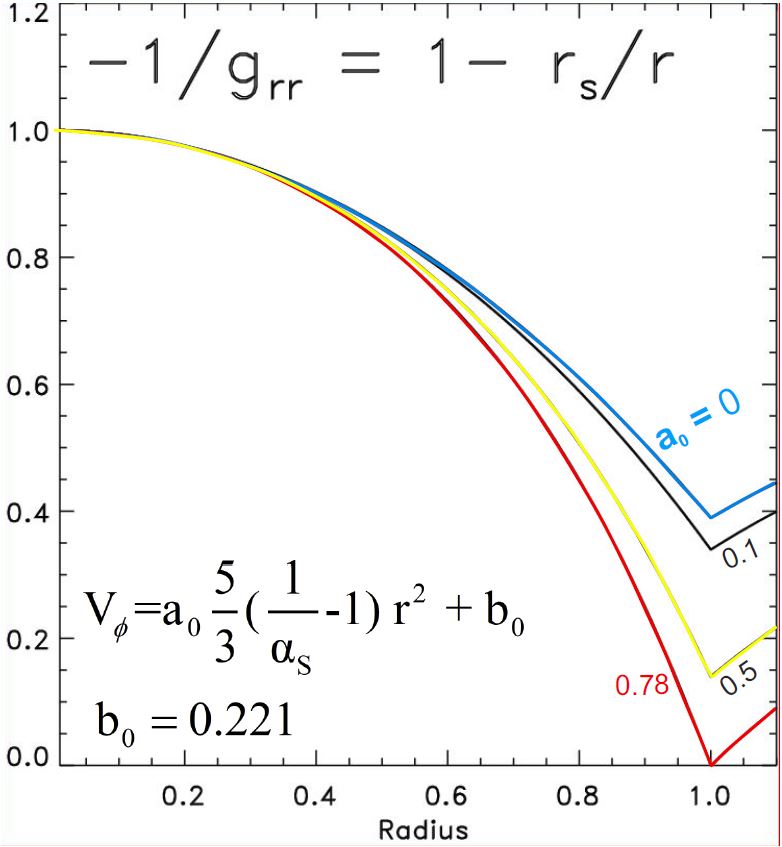}
}
\caption{\small In the top panel we show the radial distributions of the metric coefficients $g_{rr}$ and $g_{tt}$
inside a NS (; $P_L=\calK \rho^\gamma$ and $P_\phi =0$) and inside a DEO (; $P_L= const.$ and $P_\phi =-V_\phi$).
Obviously,  normal models of NSs  have larger radii and considerably less compact than their DEO-counterparts, which
can be inferred from the very limited spacial variations of $g_{rr}$ and $ g_{tt}.$   In the lower panel, the compactness of  a typical DEO,
expressed in terms of  -$1/g_{rr}$ is shown for different values of $a_0.$ The object turns invisible if  $V_\phi$ is calculated using
$a_0=0.78$ and $b_0=0.221.$
}  \label{DarEnergy_grr}
\end{figure}

\begin{figure}
\centering {\hspace*{-0.75cm}
\includegraphics*[angle=-0, width=8.15cm]{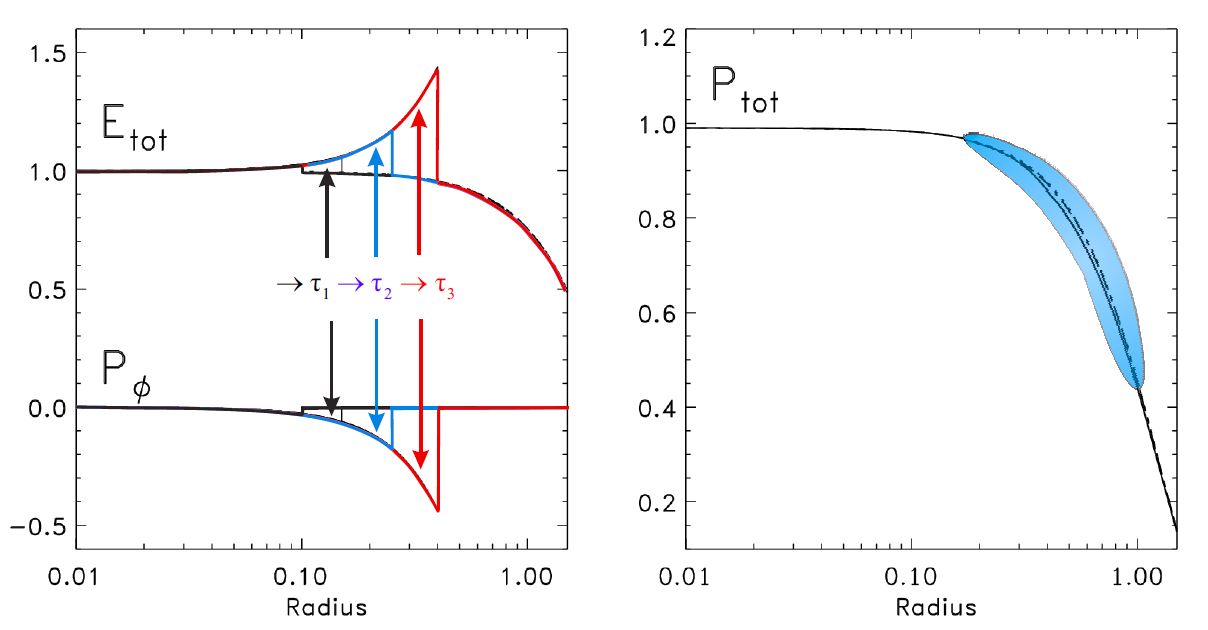}\\
}
\caption{\small The profiles of the total energy density $E_{tot},$ the  pressure $P_\phi$ induced by  $\calH$
and the combined pressure $P_{tot}$ versus radius  are shown for  different evolutionary epochs $ \tau_1 < \tau_2 < \tau_3.$
Inside $r_f$:  $P_L =0$ and $P_\phi= -V_\phi,$ whereas outside $r_f$: $P_L=\calK \rho ^\gamma$ and $P_\phi =0.$
 In each epoch, the object has an INQSF-core  overlayed by a shell of normal compressible  matter obeying a polytropic EOS.
 Obviously, the object appear to comfortably adjust itself to the mass-redistribution inside $r_f,$ where matter is converted into INQSF.
}  \label{DarEnergy_grr}
\end{figure}

\begin{figure}
\centering {
\includegraphics*[angle=-0, width=8.15cm]{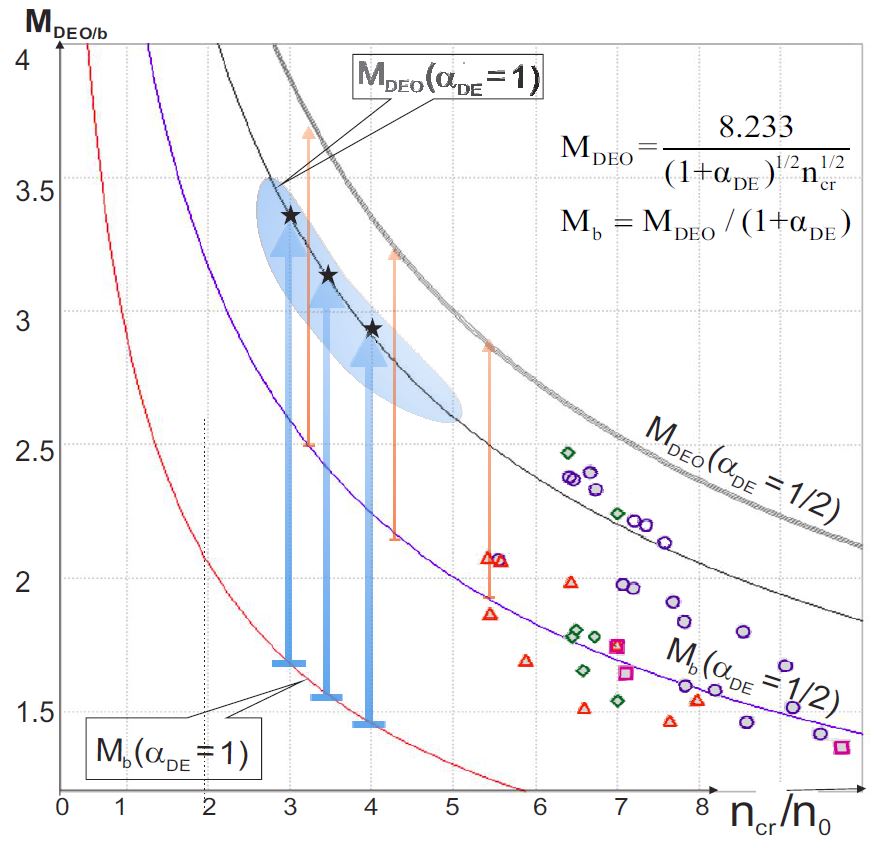}\\
}
\caption{\small   Upper mass limit of DEOs versus critical density $n_{cr}$ (in units of $n_0$) is shown. The $\calH-$baryon interaction is set
 to occur at $n_{cr},$ which in turn  provokes  the phase transition into the INQSF state. The most probable  mass-regime of  DEOs
is marked here as a blue region. Accordingly, the progenitor of a  DEO with $3.36\,\MSun$   should be a  NS of $1.68\,\MSun,$ provided it has an initial compactness $\alpha_S = 1/(1+\alpha_{DE})= 1/2$ and  $n_{cr} = 3\,n_0.$ Similarly, a Hulse-Taylor type pulsar would end as a DEO of $2.91\,\MSun,$ if  its initial compactness is $\alpha_S = 1/2$  and  if $n_{cr} = 4\,n_0.$
On the other hand, moderate and massive NSs with initial compactness $\alpha_S \geq 2/3,$ i.e., $ \alpha_{DE} \leq 1/3,$ need less dark energy to become  invisible DEOs, but  require unreasonably high $n_{cr}$ for the onset of $\calH-$matter interaction. NSs  falling in this category are to be compared
with  the colored small cycles and triangles, which show the approximate locations of various NS-models as depicted in Fig.  (4) of Lattimer \& Prakash (2011).
}  \label{NS_Vortices}
\end{figure}
\begin{figure}
\centering {\hspace*{-0.75cm}
\includegraphics*[angle=-0, width=8.15cm]{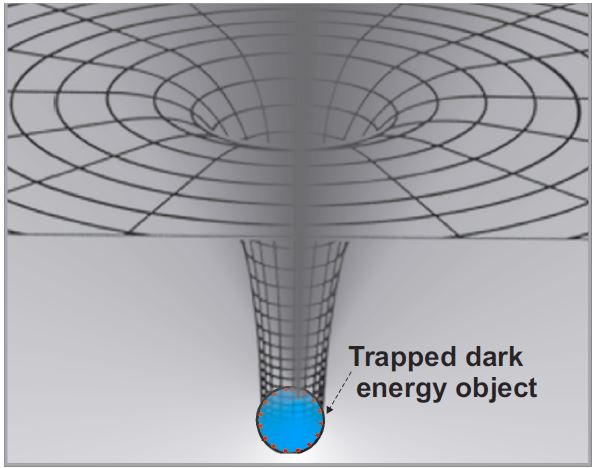}\\
}
\caption{\small A schematic description of a  DEO, inside which spacetime  is fairly flat,
but becomes extra-ordinary curved across their surfaces.
As the  binding energy inside DEOs vanishes, they are astoundingly similar to ultra-giant hadrons
trapped inside a strongly curved spacetime, which render them invisible.
}  \label{DarEnergy_grr}
\end{figure}
 \section{Results \& discussions}

 The here-presented model of DEOs is motivated by  the following three unresolved theoretical and observational problems
 in the astrophysics of NSs:
  \bit
  \item  Why  neither NSs nor BHs have ever been observed in the mass-range 2 - 5$\,\MSun.$
  \item  Most sophisticated EOS used to model the internal structure of NSs are based on central densities that are far beyond the
            nuclear density: a density regime of great uncertainty.
  \item  How NSs end their life in an ever expanding universe and whether  there is a hidden connection between the missing  massive NSs
             and dark matter on the one hand and with dark energy in the universe on the other hand.
  \eit

  In this paper we argue that the formation of DEOs may provide answers to these unresolved problems.
  This scenario could be summarized as follows:
 \ben
 \item  The very central regions of NSs are made of superfluid nuclear matter and that these would
  be the first to be evacuated from vortex lines and all other removable energies that do not contribute significantly
           to the pressure. The  nuclear fluid here is governed by the stiff EOS: $P=\calE = a_0\,n^2.$
 \item  In order to escape collapse into a BH with $M< 5~\MSun,$ the chemical potential $\mu$ in the very central regions cannot grow indefinitely, and
           it must terminate at a certain critical value, $n_{cr},$  where  the fluid is set to undergo a phase transition. \\
           Based on minimum energy consideration, a crossover phase transition into an incompressible quark-superfluid
           has been shown to be energetically a favorable transition.
\item  We have shown that in the presence of a universal scalar field $\calH,$ the injected dark energy is capable of  provoking
   a phase transition into INQSF  that roughly occurs at $ n \approx 3 ~n_0.$ The action of the injected energy  is equivalent to
   generating a gluon-like field, or  enhance the available gluon-field through forming a global energy bag in the continuum, inside which quarks move almost free in line with the asymptotic freedom of quantum chromodynamics.

\item We have shown that the transition front creeps from inside-to-outside on the scale of Gyrs, forming a sea of
quarks behind the front.  Indeed,   the very slow outwards propagation of $r_f$  grants the NS ample of time to stably react to all possible conditions
  associated with the phase transition, including a global re-distribution of mass inside $r_f$ (Fig. 6).

\item We have shown that an interaction potential of the type $V_\phi = a_0 r^2 + b_0$ is capable of maximizing the compactness of the object (Fig. 5).
\item  Having $r_f$  reached the surface of the NSs, these object become  DEOs. Their interiors are made solely of  INQSFs with constant chemical potential. The spacetime inside DEOs has been identified to be fairly flat, whereas it promptly becomes extra-ordinary curved across their surfaces (Fig. 5).
     Inside DEOs, the nuclear fluid has a vanishing binding energy and therefore mimicking the configuration of  an ultra-giant hadron
      trapped in a strongly curved spacetime.

\item According to the here-presented scenario, all visible pulsars and NSs must contain incompressible quark-superfluid cores supported and confined
by a dark energy component which is  induced by a scalar field of universal origin. The gravitational significance of the injected dark energy
in these cores depends strongly on their evolutionary phase and in particular on their  ages and initial compactness.
Accordingly, young NSs should be less massive than old ones, and  the very old NSs should  turn invisible by now.\\
To quantify the mass-enhancement by $\calH,$ let $M_b$ be the mass of the NS at its birth
 and $ M_\phi $ being the mass enhancement due to $\calH$. Requiring $R_\star > R_S,$ then
the following inequality holds:
\beq
(1 + \alpha_{DE}) \leq (\DD{3 \rho_{cr}}{32 \pi })^{1/3} \DD{c^2}{G M^{2/3}_b},
\eeq
or equivalently,
\beq
1\leq \DD{E_{tot}}{E_b}\leq   2.06 \DD{\rho^{1/3}_{15}}{M^{2/3}_{b/1.44}},
\eeq
 where $\alpha_{DE} \doteq M_\phi/M_b.$  $E_{tot},~\rho_{15},~M_{1.44}$ denote the total energy, the density in units of $10^{15}$ g/cc
 and the baryoinc mass of the NS in units of   $1.44\,M_\odot,$ respectively.\\
  Thus, NSs are born with ${E_{tot}}= {E_b},$   and by interacting with $\cal{H},$  they become more massive and more compact to finally  reach
  $R_\star = R_S + \epsilon$   at the end of their luminous phase, which would last for approximately $10^9$ yr or less, depending on their initial
  compactness. Thus, a NS  with initial compactness $\alpha_s=1/2 $  will have to double its mass to become a DEO (Fig. 4 and Fig. 7).\\
  According to the present scenario, the Hulse-Taylor pulsar should have an INQSF core, though the dark energy component is
  gravitationally insignificant due to its young age,
  and therefore the size of its INQSF-core must be still small. Assuming the baryon mass of the pulsar to remain constant as it evolve on the cosmological time scale, then
  the pulsar  will turn into invisible-DEO in roughly one Gyr. This would imply that the onset of $\calH-$baryon interaction should occur at roughly four times the nuclear density, which   is in the range of the here-predicated critical density (Fig. 3).
  On the other hand, the extra-mass resembles the lower energy limit required for deconfining  the sea of quarks,  i.e., the energy needed for
  generating a see of quark anti-quark pairs. Similar to quarks in hadrons, the sea of quarks inside DEOs can never be observed as free objects in the sky.\\

 \een

 Recalling that the effective potential of the gluon-field inside individual hadrons is on the average  predicted to increases with radius as $r^{\Gamma(>1)}$ and that
  the  spatial variation of  the coefficient $g_{rr}$ of the Schwarzschild metric on comparable length scales is negligibly small
  ($dg_{rr}/dl \ll 10^{-19}$),  we conclude that gluon-fields do not accept stratification by gravitational fields.\\
  Therefore  as $E_\phi $ in the present DEO-models is dominant and increases with radius,  the
  sea of quarks inside DEOs is in a purely incompressible state and cannot accept stratification (see $g_{tt}$ in Fig. 5).
    In such gravitationally bounded incompressible fluid-configurations, not only that $\mu = \calE=const.,$ but the classical repulsive pressure
    $P_L (\doteq n_b^2 \DD{\D}{\D n}(\DD{\calE}{n_b}))$ must vanish also and should be replaced by a non-local pressure, $P_{NL},$
    in order to avoid the formation of BHs with $M< 5~\MSun.$


            Unlike EOSs in compressible normal plasmas,  classical EOSs in incompressible superfluids are non-local. In the latter case,
            constructing a communicator that merely depends on local exchange of information generally  would not be sufficient for efficiently coupling
            different/remote parts of  the fluid in a physically consistent manner.  A relevant example is the solution of the TOV-equation for classical
            incompressible fluids $(\calE =const.)$.
            In this case, the pressure depends, not only on the global compactness of the object, but it becomes even acausal whenever the global compactness is enhanced.\\
            This is similar to the case when solving the incompressible Navier-Stokes equations, where an additional Laplacian operator for  describing the spatial variation             of a non-local scalar field  is constructed to generate  a pseudo-pressure (; actually a Lagrangian multiplier), which, again, does not respect causality \cite[][]{Hujeirat2009}.



Indeed, DEOs made of  incompressible quark-superfluids  would be stable also against mass-enhancement from outside. Let a certain amount of
baryonic matter, $\delta \calM_b,$   be added to the object from outside. Then the relative increase of $R_\star$ compared to
             $R_S$ scales as:  $\DD{\delta R_\star}{\delta R_S} \simeq \DD{\rho_{cr}}{\tilde{\rho}_{new}},$  where
            $\tilde{\rho}_{new}$ is the average density of the newly settled matter. Unless  $\tilde{\rho}_{new} > \rho_{cr},$
            which is forbidden under normal astrophysical conditions,  the star would react stably. However, in the case of super-Eddington accretion or merger,
            the newly settled  matter  must first decelerate, compressed and subsequently becomes virially hot, giving rise therefore to
             $\tilde{\rho}_{new} \ll  \rho_{cr}.$  On the other hand, such events would lower the confinement stress at  the surface
             and would turn the quantum jump of the energy density  at $R_\star,$ which falls abruptly  from  approximately $\calE \approx10^{36}$ erg/cc at $R_\star$ down to zero outside it,  into an extra-ordinary steep pressure gradient in  the continuum.
             While such  actions would smooth the strong curvature of spacetime across $R_\star,$  they
              would enable DEOs to eject quark matter into space with ultra-relativistic speeds, which is forbidden.
              Nonetheless, even if this would occur instantly, then the corresponding time scale $\tau_d$ would be
            of order $\Lambda_j/c,$ where $\Lambda_j$ is the jump width in centimeters and c is the speed of light. Relating $\Lambda_j$ to the average spacing between
            two arbitrary particles $(\sim n^{-1/3}),$ this yields  $\tau_d \approx 10^{-24}\,$s, which is  many orders of magnitude shorter
            than any known thermal relaxation time scale between arbitrary luminous particles.\\

            Although electromagnetic activities and jets have not been observed in dark matter halos, they are typical events for systems
            containing  black holes. Recalling that supermasive GBECs are dynamically unstable \cite[][]{Hujeirat2012},
             our results here address the following two possibilities:
             \bit
            \item  If the onset of $\calH$-baryon interaction indeed occurs at $n_{cr},$ then the majority of the  first generation of stars and the massive
            stars formed in the subsequent early epochs must have ended as  pulsars and NSs, rather than collapsing into
             stellar BHs with $M \leq 5\times \MSun$. In this case,  dark matter halos most likely should be  DEO-rich clusters. These clusters must have been extraordinary luminous
                in the early universe, but became inactive and dark after the nuclear matter in the interiors of NSs converted into the INQSF-phase, subsequently sweeping away  all sorts of  luminous matter in their surroundings due to their inability to accrete normal matter. The enormous surface stress
                confining the sea of quarks in the interiors of DEOs render their surfaces impenetrable for normal matter, hence these objects
                 behave as non-interacting objects.
            \item The average repulsive forces governing clusters of DEOs most likely would enforce approaching luminous matter to deviate from face-to-face collisions and  therefore stay inactive, though n-body and SPH-numerical calculations are needed here to verify this argument.
            \eit

             Finally we note that, similar to the gluon field confining and governing the dynamics of almost massless quarks
             in hadrons,  the $\calH-$induced energy enhancement of  the gluon-like field in DEOs cannot surpass the limit,
             beyond which they collapse to form BHs with $M< 5~\MSun.$ Moreover, the enclosed dark energy injected via
             $V_\phi$ scales as $r^5$: this outlines an upper limit
             for the increase of confining energy with radius in DEOs, beyond which they undergo a self-collapse.
             However, whether this limit applies for the potential of the gluon field inside hadrons is not clear at the
              moment and demands further investigations.

           In a subsequent article, we discuss the compatibility and physically consistency of the here-presented
           internal structures of DEOs with  the bi-metric formulation  of spacetime in general relativity proposed by \cite{Rosen1977}.\\

\textbf{Acknowledgment} The author thanks Johanna Stachel, Friedel Thielemann, James Lattimer, George Chapline, 
                                 Tsvi Piran,  Juergen Berges,  Jan Martin Pawlowski, Max Camenzind, Ravi Samtaney and Matthias Hempel
                                 for the very helpful comments and useful discussions on various aspects of this article.


\begin{thebibliography}{99}
\bibitem[Alpar(1984)]{Alpar1984} Alpar, M.A., Anderson, P.W., Pines, D., Shaham, J., ApJ, 278, 791, 1984
\bibitem[Baggaley \& Laurie(2014)]{Baggaley2014} Baggaley, A.W., Laurie, J., Phys. Rev. B., 89, 014504, 2014
\bibitem[Baym(1995)]{Baym1995} Baym, G., Nuclear Physics A, 233, 1995
\bibitem[Baym \& Chin (1976)]{Baym1976} Baym, G., Chin, S.A., Phys. Lett. 62B, 241, 1976
\bibitem[Baranghi(2008)]{Baranghi2008}Baranghi, C., Physica D, 237, 2195, 2008
\bibitem[Belczynski(2012)]{Belczynski2012} Belczynski, K.,  Wiktorowicz, G.,  Fryer, C., et al., ApJ, 757, 91, 2012
\bibitem[Bethke(2007)]{Bethke2007} Bethke, S., Progress in Particle and Nuclear Physics, 351, 58, 2007
Belczynski, S., Progress in Particle and Nuclear Physics, 351, 58, 2007
\bibitem[Camenzind (2007)]{Camenzind2007}  Camenzind, M., "Compact Objects in Astrophysics", Springer, 2007
\bibitem[Chapline(2014)]{Chapline2014}  Chapline, G., Barbieri, J., International Journal of Modern Physics D,  23, No. 3, 1450025, 2014
\bibitem[Dix(2014)]{Dix2014} Dix, O.M., Zieve, R.J., Physical Review B, 144511, 90, 2014
\bibitem[Espinoza(2011)]{Espinoza2011} Espinoza, C.M.,  Lyne, A.G., Stappers, B.W., Kramer, C., MNRAS, 414, 1679, 2011
\bibitem[Hempel et al. (2011)]{Hempel2011}  Hampel, M., Fischer, T., et al., ApJ, 748, 70, 2012
\bibitem[Haensel et al. (2007)]{Haensel2007}  Haensel, P., Potekhin, A.Y. \& Yakovlev, D.G., "Neutron stars 1", Springer, 2007
\bibitem[Hujeirat \& Thielemann(2009)]{Hujeirat2009} Hujeirat, A.A., Thielemann, F-K., MNRAS, 400, 903, 2009
\bibitem[Hujeirat(2012)]{Hujeirat2012} Hujeirat, A.A., MNRAS, 423.2893, 2012
\bibitem[Hujeirat(2016)]{Hujeirat2016}  Hujeirat, A.A., in preparation, 2016
\bibitem[Kalogera(1996)]{Kalogera1996} Kalogera, V., Baym, G.,  APJ, 470, L61, 1996
\bibitem[Lattimer(2011)]{Lattimer2011} Lattimer, J.M.,  Prakash, M., in "From Nuclei To Stars", Ed.  S. Lee, World Scientific Publishing, Singapur, 2011
\bibitem[Link(2012)]{Link2012}  Link, B.,   MNRAS, 422, 1640, 2012
\bibitem[Rosen(1977)]{Rosen1977} Rosen, N., ApJ, 211, 357, 1977
\end{thebibliography}
\end{document}